\documentclass[twocolumn,preprintnumbers,superscriptaddress,nofootinbib,aps,prd,floatfix]{revtex4}

\usepackage{footmisc}
\usepackage{amsmath}
\usepackage{wasysym}
\usepackage{graphicx}
\usepackage{color,array,subfigure}
\usepackage{comment}

\newcommand{\be}{\begin{eqnarray}}
\newcommand{\ee}{\end{eqnarray}}

\newcommand{\bee}{\begin{eqnarray}}
\newcommand{\eee}{\end{eqnarray}}
\newcommand{\beeq}{\begin{equation}}
\newcommand{\eeeq}{\end{equation}}

\numberwithin{equation}{section}

\begin{document}

\title{LHC Signatures Of Scalar Dark Energy}

\begin{abstract}
Scalar dark energy fields that couple to the Standard Model can give rise to observable signatures at the LHC. In this work we show that $t\bar t+$missing energy and mono-jet searches are suitable probes in the limit where the dark energy scalar is stable on collider distances.
We  discuss the prospects of distinguishing the dark energy character of new physics signals from dark matter signatures and the possibility of probing the self-interactions of the dark energy sector.
\end{abstract}

\author{Philippe Brax} \email{philippe.brax@cea.fr}
\affiliation{Institut de Physique Th\'{e}orique, Universit\'e Paris-Saclay, CEA, CNRS, F-91191 Gif/Yvette Cedex, France\\[0.1cm]}
\author{Clare Burrage} \email{clare.burrage@nottingham.ac.uk}
\affiliation{School of Physics and Astronomy, University of Nottingham, Nottingham, NG7 2RD, United Kingdom\\[0.1cm]}
\author{Christoph Englert} \email{christoph.englert@glasgow.ac.uk}
\affiliation{SUPA, School of Physics and Astronomy, University of
  Glasgow,\\Glasgow G12 8QQ, UK \\[0.1cm]}
\author{Michael Spannowsky} \email{michael.spannowsky@durham.ac.uk}
\affiliation{Institute for Particle Physics Phenomenology, Department
  of Physics,\\Durham University, Durham DH1 3LE, UK\\[0.1cm]}

\preprint{IPPP/16/31, DCPT/16/62}

\maketitle

\section{Introduction}
\label{sec:intro}
The expansion of the universe is currently accelerating, and yet we have no compelling explanation of why this is happening unless we are prepared to accept the extraordinary degree of fine tuning associated with the introduction of a cosmological constant.  Attempts to further our understanding typically introduce new scalar fields either explicitly as quintessence, or implicitly through a modification of the gravitational sector~\cite{Copeland:2006wr,Clifton:2011jh,Joyce:2014kja}.  It is therefore crucial for cosmology to understand what theoretical properties these scalar fields could have, and to   constrain them  experimentally; whilst remaining agnostic about the complete solution of the cosmological constant problem and the source of the acceleration of the expansion of the universe.

A lot of attention has been recently focused on the Horndeski theories, which are the most general theories describing one scalar field coupled to gravity~\cite{Horndeski:1974wa}, that have second order equations of motion.  These theories were first written down by Horndeski, and later independently rediscovered by Deffayet, Gao, Steer and Zahariade~\cite{Deffayet:2011gz}.  Insisting on second order equations of motion guarantees the absence of ghost degrees of freedom, although it has also been realised that if additional constraints are present this condition can be relaxed and the theories extended to the so called `beyond-Horndeski' theories~\cite{Gleyzes:2014dya}. The Horndeski theories  provide a complete description of the possible effects of a new scalar degree of freedom uniformly coupled to matter, and  constraining these theories is an important target for upcoming large scale cosmological surveys including Euclid~\cite{Amendola:2012ys}.

 Such a dark energy scalar field may arise as part of a solution to the cosmological constant problem; the question of why the vacuum fluctuations of standard model fields do not generate a large effective cosmological constant. Any solution to this problem must therefore interact to both the gravitational and matter fields.  Therefore, bar any otherwise compelling reason, we expect that the dark energy scalar will couple to matter~\cite{Joyce:2014kja}.  This is potentially problematic, because light scalar fields coupled to matter mediate fifth forces.  The stringent experimental constraints on the existence of such forces can be avoided, either by imposing a shift symmetry which forbids Yukawa type interactions with the scalar, or by making the theory non-linear and thereby allowing the properties of the fifth force to vary depending on the environment, an effect known as screening \cite{Joyce:2014kja}.

The energy scales relevant to dark energy are the (reduced) Planck mass $M_P = 2.4 \times 10^{18}~\text{GeV}$ controlling the strength of gravitational effects, and the Hubble scale today $H_0= 1.5 \times 10^{-42}~\text{GeV}$ which sets the coherence scale for dark energy effects.  The vast hierarchy between these two scales is the source of the cosmological constant problem, which we do not address here, but it also allows us to build a vast array of intermediate scales by taking different combinations of the Planck mass and the Hubble scale. For example Dvali-Gabadadze-Porrati (DGP) and Galileon models have higher mass dimension scalar operators suppressed by the scale $ (M_P H_0^2)^{1/3} \sim 10^{-22}~\text{GeV} $~\cite{Dvali:2000hr,Nicolis:2004qq,Nicolis:2008in}. The invention and widespread adoption of screening mechanisms to dynamically suppress fifth forces in intra-solar-system searches~\cite{Joyce:2014kja} also means that experimental bounds can be met without the energy scale controlling the strength of the coupling of the scalar to matter being forced to lie above the Planck scale.

As a result we should ask whether it is possible to detect the Horndeski model of dark energy  on terrestrial scales.  Constraints from laboratory experiments will provide important information, complementary to that obtained from cosmological surveys, and allows us to test theories of dark energy over the widest possible range of distance and energy scales.

The LHC probes our understanding of physics at unprecedented energies and under controlled and reproducible conditions. A large variety of particles, including ones with heavy masses, are being produced beyond threshold, resulting in potentially sizeable interactions of new scalars that couple to standard model (SM) particles via the energy-momentum tensor. In doing so, the LHC creates a controlled and non-static environment in the sense that large momentum transfers of physical systems are probed with sufficient accuracy and statistics. Since interactions of a scalar dark energy candidate with the SM sector and itself often involve derivative couplings, we can expect the high momentum transfer events at the LHC to provide an excellent strategy to constrain such realisations of dark energy.

In this work we will survey the modified phenomenology of LHC processes that are particularly motivated as probes of dark energy interactions. Before we discuss these processes in detail in Sec.~\ref{sec:colla}, to make this work self-contained, we survey effective dark energy models in Sec.~\ref{sec:eff} to introduce the relevant dark energy effective theory (EFT) interactions. Although different in fundamental aspects, dark energy phenomenology at the LHC shares certain aspects with searches for dark matter at colliders. The potential to pin down the dark energy character of a potential new physics signal due to different a priori phenomenology and the  expected non-linear self-interactions will be discussed in Secs.~\ref{sec:darkmatter} and~\ref{sec:collb}. We give our conclusions and an outlook in Sec.~\ref{sec:conclusions}.

\section{Effective Models for Dark Energy}
\label{sec:eff}

We consider the effective role that dark energy could play in collider experiments. Our starting point is a dark energy scalar  field $\phi$ with a comparably small
mass compared to particle physics scales. We will differentiate in what follows between theories which respect the shift symmetry $\phi \rightarrow \phi + c$, and those that break it.

We assume that  $\phi$ couples  to matter universally, in such a way that matter fields move on geodesics of the  Jordan frame metric
\begin{equation}
g_{\mu\nu}= A^2(\phi,X)\tilde{g}_{\mu\nu} + B(\phi,X) \partial_\mu\phi\partial_\nu\phi
\end{equation}
where $X=\frac{1}{2} \eta^{\mu\nu} \partial_\mu \partial_\nu \phi$. 
We assume that in the collider environment the Einstein frame metric is simply the Minkowski metric $\tilde{g}_{\mu\nu}=\eta_{\mu\nu}$, which is certainly a reasonable assumption on earth where Newton's potential is very small.

Expanding the coupling functions $A$ and $B$  in powers of $\partial_\mu\phi\partial_\nu \phi$ gives a tower of characteristic interactions. In particular we can write
\begin{equation}
A(\phi,X)=\sum_n \frac{a_n(\phi/{M})}{M^{4N}} X^n
\end{equation}
 and
\begin{equation}
B(\phi,X)=\sum_n \frac{b_n({\phi}/{M})}{M^{4N}} X^n
\end{equation}
where $a_n$ and $b_n$ are dimensionless, and become constant and independent of $\phi$ when the shift symmetry is imposed.

\subsection{Shift symmetric theories}
\subsubsection{Coupling to matter}
Assuming that the model is shift symmetric under $\phi \to \phi +c$
the lowest order interactions between the scalar and the Standard Model are through the Lagrangian terms
\begin{equation}
\label{eq:c1}
\mathcal{L}_1 = \frac{\partial_{\mu}\phi\partial^{\mu} \phi}{M^4}T^{\nu}_{\;\nu}
\end{equation}
corresponding to a direct conformal coupling with constant $a_1$,
and the disformal coupling
\begin{equation}
\label{eq:c2}
\mathcal{L}_2 = \frac{\partial_{\mu}\phi\partial_{\nu} \phi}{M^4}T^{\mu\nu}
\end{equation}
associated with a constant $b_1$.
Here $T_{\mu\nu}$ is the energy momentum tensor of all of the standard model fields. Note that no coupling between the scalar and photons arises from $\mathcal{L}_1$.

Higher order operators can have the following forms:
\begin{equation}
\mathcal{L}_{3,n} = \left(\frac{\partial_{\mu}\phi\partial^{\mu} \phi}{M^4}\right)^nT^{\nu}_{\;\nu}
\end{equation}
coming from a constant $a_n$ and
\begin{equation}
\mathcal{L}_{4,n} = \left(\frac{\partial_{\alpha}\phi\partial^{\alpha} \phi}{M^4}\right)^n\frac{\partial_{\mu}\phi\partial_{\nu} \phi}{M^4}T^{\mu\nu}
\end{equation}
from  the cross term between a constant $b_1$ and a constant $a_n$. Finally we can have higher order terms of the form
\begin{multline}
\mathcal{L}_{5,n-1} = \frac{1}{M^{4n}}\partial_{\alpha_1}\phi\partial_{\beta_1} \phi \ldots \partial_{\alpha_n}\phi\partial_{\beta_n} \phi\\
 \frac{2^{n-1}}{\sqrt{-g}}\frac{\partial^{n-1}(\sqrt{-g}T^{\alpha_1\beta_1})}{\partial g_{\alpha_2 \beta_2}\ldots \partial g_{\alpha_n \beta_n}}
\end{multline}
where $n$ is a positive integer.  The form of $\mathcal{L}_5$ is derived in~\cite{Brax:2014vva}.

\subsubsection{Kinetic terms}

Possible kinetic terms for the scalar fall into two classes.  The first, known as $P(X)$, have the form
\begin{equation}
\mathcal{L}_{6,n} = \frac{(\partial_\mu\phi \partial^{\mu} \phi)^n}{M^{4(n-1)}}
\end{equation}
for positive integer $m$.  A particular series of such operators, $\mathcal{L} = M^4 \sqrt{1+\partial_{\mu}\phi\partial^{\mu}\phi /M^4}$, arises in DBI theories~\cite{Silverstein:2003hf}, where the theory possesses an additional (non-linearly realised) symmetry which encodes 5d Lorentz invariance when $\phi$ is viewed as determining the position of a D3 brane in 5d Minkowski space. This extra symmetry allows the field to acquire large gradients while remaining in the regime of validity of the EFT.

The second class of kinetic terms are known as the Galileons, and contain terms with more than one derivative per field.  Around flat space they are invariant (up to total derivatives) under the symmetry $\phi \rightarrow \phi + c + b_{\mu}x^{\mu}$ for constant $c$ and $b_{\mu}$.  There are five Galileon operators, but one is the tadpole and one is the canonical kinetic term, so there are only three more terms we need to consider:
\begin{equation}
\mathcal{L}_7 = \frac{1}{M^3} \partial_{\mu}\phi\partial^{\mu}\phi \Box \phi
\end{equation}
\begin{equation}
\mathcal{L}_8 = \frac{1}{M^6} \partial_{\mu} \phi\partial^{\mu}\phi\left[2(\Box \phi)^2 -2 D_\alpha D_\beta \phi D^\beta D^\alpha \phi\right]
\end{equation}
\begin{multline}
\mathcal{L}_9 = \frac{1}{M^9} \partial_{\mu} \phi\partial^{\mu}\phi\left[(\Box\phi)^3 -3(\Box\phi)D_\alpha D_\beta \phi D^\beta D^\alpha \phi \right. \\
\left. + 2 D_\alpha D^\beta \phi D_\beta D^\gamma\phi D_\gamma D^\alpha\phi\right]
\end{multline}

It has been shown for both $P(X)$ and Galileon theories, that while the scale $M$ in these operators is the strong coupling scale controlling self interactions of the scalar, the effective field theory description remains valid up to a higher cut-off scale \cite{deRham:2014wfa}.

\subsection{Breaking the shift symmetry}
This set of operators can be extended further if the shift symmetry is broken and terms depending   on the undifferentiated scalar field  are allowed. A scalar theory with a softly broken shift symmetry can still be cosmologically relevant, however it suffers from issues of fine tuning, because it is necessary to keep the mass of the field light enough that it has a cosmologically relevant Compton wavelength. If we take $n$ to be a positive integer and $N$ is the energy scale that enters with $\phi$,   then each of the operators $\mathcal{L}_1$ - $\mathcal{L}_9$ can be pre-multiplied by a factor of $(\phi/N)^n$.  There are two other possibilities which depend only on $\phi$. Firstly the coupling to matter can take the form
\begin{equation}
\mathcal{L}_{10,n} = \left(\frac{\phi}{N}\right)^n T^{\mu}_{\;\mu}.
\end{equation}
For a canonical scalar with an $m^2\phi^2$ potential this form of the coupling is extremely well constrained by fifth force searches~\cite{Adelberger:2003zx}.  But in more complex and non-linear models collider bounds can still provide new information~\cite{Brax:2009aw,Brax:2009ey}.

Secondly we can include potential terms for the scalar
\begin{equation}
\mathcal{L}_{11,n} = \frac{\phi^n}{N^{n-4}},
\end{equation}
where $n$ can be either positive or negative. When $n=1$ this is a tadpole that, as mentioned above, we ignore.  When $n=2$ this is a mass term for the scalar, which it will be helpful to consider separately in what follows.

\subsection{Ghosts}
The above list clearly does not include all possible operators that depend on $\phi$ and its derivatives.  However the remaining terms will introduce ghost degrees of freedom, that is fields with negative norms or wrong sign kinetic terms, leading to instabilities and a violation of unitarity.  These terms have the schematic form
\begin{equation}
\mathcal{L}_{12,m,n}= \frac{\partial^m\phi^n}{M^{m+n-4}}
\end{equation}
with $m>n>1$ and the derivatives are contracted in a Lorentz invariant way, and can be included in our effective field theory as long as they handled with care, as  the instabilities introduced by the ghost only appear at the scale $M$ assumed to lie close to the cut off of the theory, at which our effective treatment breaks down.

The exception to this are the so-called beyond Horndeski theories which contain non trivial constraints that remove the ghost degrees of freedom introduced by these operators. The nature of these constraints means that they are difficult to study on an operator by operator basis.

\section{Collider Phenomenology}
\label{sec:coll}
We now consider the collider phenomenology of the operators introduced above by writing
\begin{equation}
{\cal{L}}_{\text{BSM}}={\cal{L}}_{\text{SM}}+\sum_{i}C_i {\cal{L}}_i + {1\over 2}m_\phi^2\phi^2	\,,
\end{equation}
with Wilson coefficients $C_i$ and we limit ourselves to the lowest non-trivial orders in each operator series. The production cross sections of a given multiplicity of $\phi$ scalars depends on the ratio $\sim C_i^2/M^{2r}$, where $r$ is the characteristic scaling of the operators listed above. We will choose $C_i=1$ to report constraints solely expressed by the scale $M$, but it should be understood that $C_i\neq 1$ are possible choices, too. As already mentioned, we focus on light values of $m_\phi$ in comparison to typical collider scales; we adopt $m_\phi=0.1~\text{GeV}$ as our benchmark in the following.

Out of the operators of the previous section, $\mathcal{L}_{10}$ is special as it enables the prompt decay of $\phi$ into SM fields if sufficient phase space is available. This changes the LHC phenomenology dramatically, also because single $\phi$ production becomes available, only suppressed by $\sim N^{-1}$, thus giving rise to a possibly dominant contribution. The mass $m_\phi$ becomes a crucial parameter in this case and there LHC analysis strategies will be fundamentally different from the situation when $\phi$ is stable on collider length scales. We will not discuss this possibility in detail at this stage but provide a qualitative discussion in Sec.~\ref{sec:shift} and leave a detailed analysis of the shift-symmetry breaking phenomenology to future work.

Not considering $\mathcal{L}_{10,11}$ for the moment, the dominant phenomenological signature is missing energy as the pair-produced scalar particles escape detection on collider scales. In a phenomenological bottom-up approach, such a signature can be attributed to a plethora of models ranging from Supersymmetry over general dark matter signatures to extra dimensions. The operators listed in the previous section, however, have a significantly modified phenomenology due to their particular derivative structure and characteristic mass suppression, in addition to their relation to the energy momentum tensor. This also provides an opportunity to address the inverse problem by directly investigating the non-linear structure of the $\phi$ interactions and their impact on LHC phenomenology. In the following, we will
	identify suitable search channels for the scenarios discussed in the previous section, extending beyond available investigations~\cite{Brax:2015hma}, specifically with the aim to distinguish the leading EFT operators ${\cal{L}}_1$ and ${\cal{L}}_2$. We will also investigate the characteristic behaviour of non-linearities and discuss the prospects to pin down the dark energy character of a missing energy signature if such an observation is made at the LHC in the future. We will then come back to broken shift symmetry operators to discuss their phenomenological impact. Throughout we use the combination of {\sc{FeynRules}}~\cite{Alloul:2013bka}, {\sc{Ufo}}~\cite{Degrande:2011ua}, and {\sc{MadGraph5}}~\cite{Alwall:2014hca} to simulate the final states.

\subsection{Dark energy signatures at the LHC}
\label{sec:colla}

\begin{figure}[!t]
  \centering
  \includegraphics[height=6.0cm]{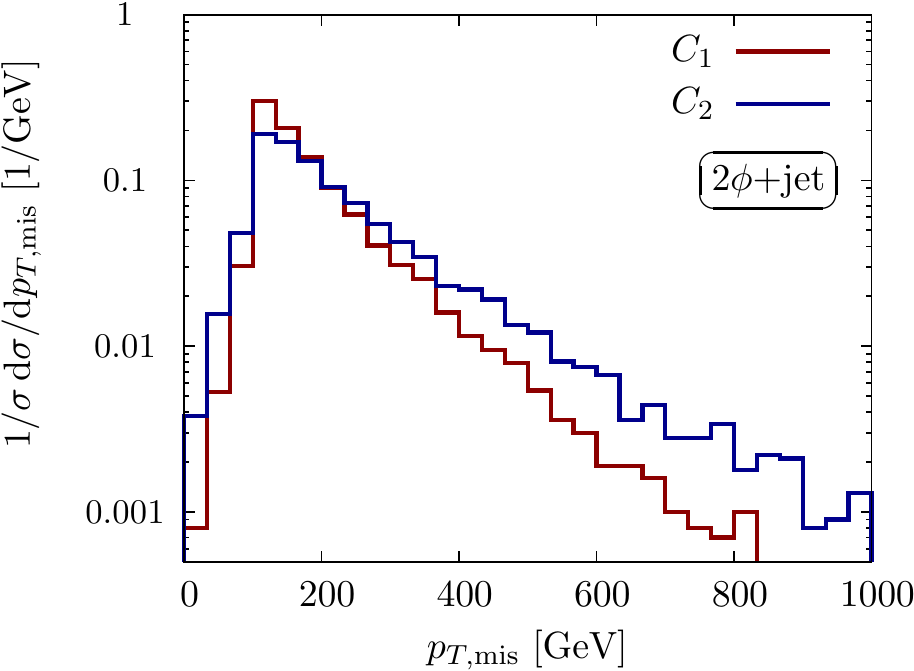}
  \caption{\label{jet:c1c2comp} Shape comparison of the jet+missing transverse momentum distribution for conformal and disformal couplings, Eqs.~\eqref{eq:c1} and \eqref{eq:c2}.}
\end{figure}

Under the assumption that $\phi$ is stable on collider scales, the dominant signature is missing energy as the visible particles recoil against the invisible and pair-produced $\phi$ bosons. There is a comprehensive catalogue of missing energy searches, mostly interpreted in a Supersymmetry or dark matter-related context. Channels that have been scrutinised recently are mono-boson production in association with missing energy (e.g.~\cite{cmsphoton,atlasphoton,cmslepton,atlaslepton,atlasdilepton}) and mono-jet searches~\cite{cmsjets,atlasjet,Chatrchyan:2013mys,Aad:2014wea}. The latter have been identified as excellent candidates to constrain disformal couplings $\sim T^{\mu\nu}\partial_\mu\phi \partial_\nu\phi$ in~\cite{Brax:2015hma} motivated by the large momentum transfers that are probed with sufficient statistics in the mono-jet signal, especially for the high missing energy selections of~\cite{cmsjets}.

Turning to the operator of Eq.~\eqref{eq:c1}, the scaling arguments of the $2\phi
$+jet signature still hold, see Fig.~\ref{jet:c1c2comp}. The crucial difference between the ${\cal{L}}_1$ and ${\cal{L}}_2$ couplings lies in the fact that the coupling to the trace of the energy momentum tensor is tantamount to coupling the $\phi$ pairs to all explicit conformal invariance-violating terms in the Standard Model, in particular to all mass terms. This results in an extremely small cross section of the mono-jet final states as the quark masses are small and the hadronic cross section receives a large contribution from massless gluons. Using {\sc{CheckMate}}~\cite{Drees:2013wra} to survey ATLAS and CMS mono-jet analyses we can only set a constraint at 95\% confidence level of\footnote{We only quote the most sensitive search region in the respective analyses.}
\begin{equation}
\label{eq:liml1j}
\parbox{0.37\textwidth}{
\begin{tabular}{c >{\hspace{0.3cm}} l >{\hspace{0.3cm}} r }
${\cal{L}}_1 $ & $M\gtrsim 75.4~\text{GeV}$ & $\text{(ATLAS~\cite{atlasjet})}$ \\[0.1cm]
$2\phi+\text{jet}$ & $M\gtrsim 66.5~\text{GeV}$ & $\text{(CMS~\cite{Chatrchyan:2013mys})}$ \\
\end{tabular}
}\,,
\end{equation}

\begin{figure}[!t]
  \centering
  \includegraphics[height=6.0cm]{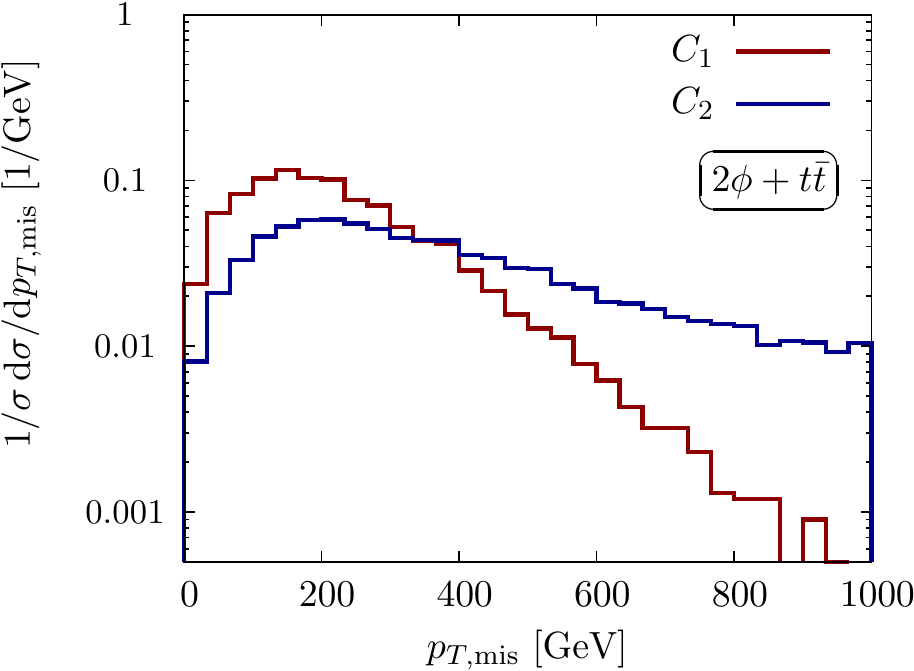}
  \caption{\label{top:c1c2compt} Shape comparison of the $t\bar t$+transverse momentum distribution in the presence of conformal and disformal couplings, Eqs.~\eqref{eq:c1} and \eqref{eq:c2}.}
\end{figure}

\begin{figure*}[!t]
  \centering
  \subfigure[\label{jet:dmcomp}]{\includegraphics[height=6.0cm]{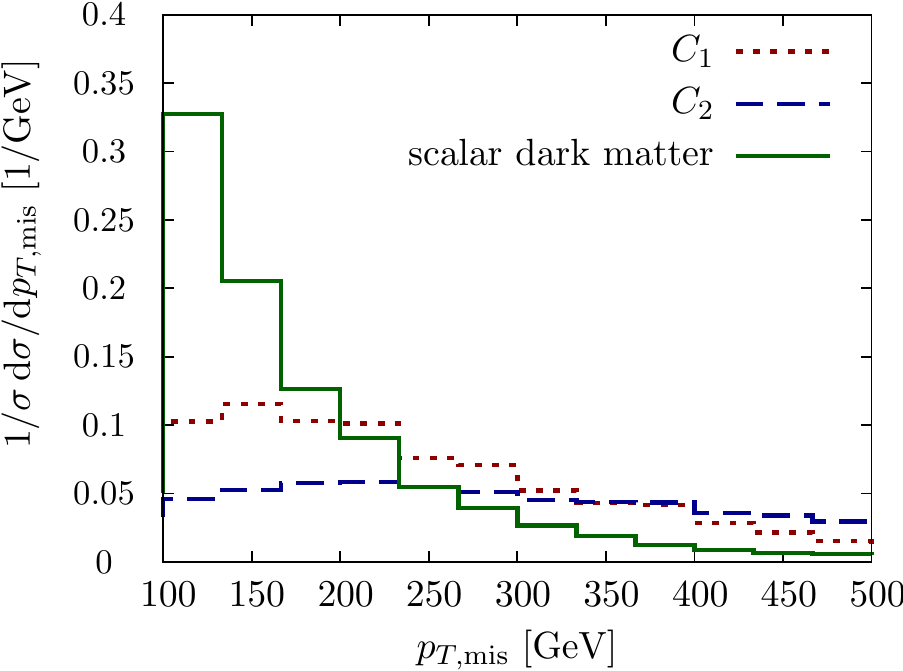}}
  \hfill
  \subfigure[\label{top:dmcomp}]{\includegraphics[height=6.0cm]{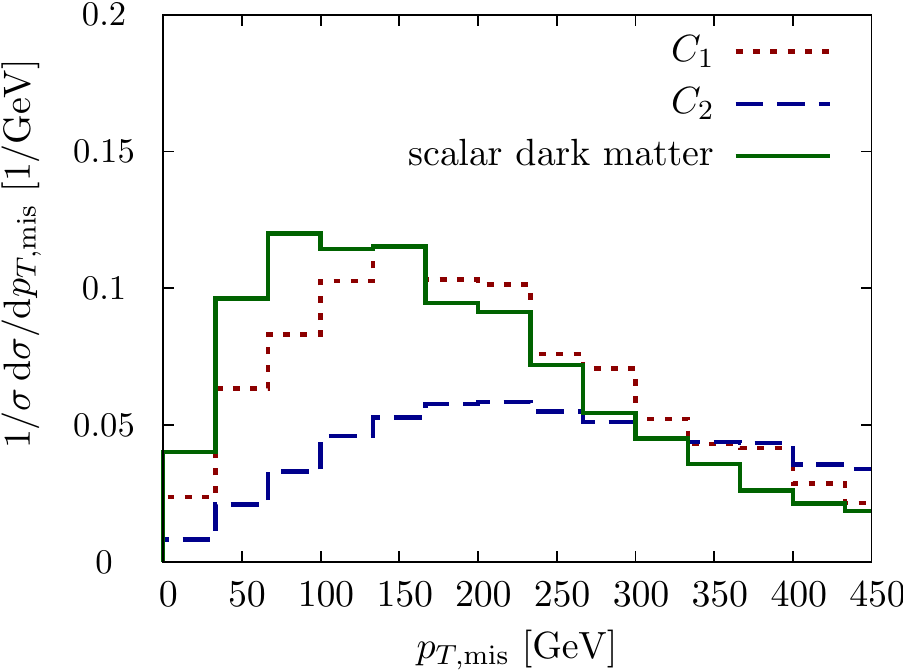}}
  \caption{\label{fig:dmcomp} Shape comparison of the dark energy scalar $p_{T,\text{mis}}$ distribution for (a) $2\phi$+jet and (b) $2\phi+t\bar t$ with a scalar mediator of mass $1~$TeV.}
\end{figure*}

The observation that ${\cal{L}}_1$ is directly related to explicit mass scales, however, directly motivates top quark production in association with missing energy. The reason for this is twofold. Firstly, the top quark is the heaviest particle in the SM, and as a consequence will have a large ${\cal{L}}_1$-mediated coupling to the dark energy scalars. Secondly, top quark pair production with a total strong interaction-dominated production cross section of around $900$~pb at 13 TeV is far more accessible than the Higgs boson, which would be motivated as a potential probe of ${\cal{L}}_1$ along the same line of arguments. Indeed, we find that $2\phi+t\bar t$ production has a significant cross section for $C_1\neq 0$ and setting more stringent limits becomes possible. We find
\begin{equation}
\label{eq:liml1t}
\parbox{0.37\textwidth}{
\begin{tabular}{c >{\hspace{0.3cm}} l >{\hspace{0.3cm}} r }
${\cal{L}}_1 $ & $M\gtrsim 237.4~\text{GeV}$ & $\text{(ATLAS~\cite{ATLAS-CONF-2013-024})}$ \\[0.1cm]
$2\phi+t\bar t$ & $M\gtrsim 192.8~\text{GeV}$ & $\text{(CMS~\cite{Chatrchyan:2013mys})}$\\
\end{tabular}
}
\end{equation}

This not only motivates $t\bar t + p_{T,\text{mis}}$ searches as probes for dark energy scalars, but in particular the combination of mono-jet and top pair$+ p_{T,\text{mis}}$ searches can provide a fine-grained picture of the phenomenology of ${\cal{L}}_{1,2}$ as we will see in the following when we study the effects of ${\cal{L}}_2$.

For the mono-jet signatures, the most constraining 8 TeV analyses yield
\begin{equation}
\label{eq:liml2j}
\parbox{0.37\textwidth}{
\begin{tabular}{c >{\hspace{0.3cm}} l >{\hspace{0.3cm}} r }
${\cal{L}}_2 $ & $M\gtrsim 693.9~\text{GeV}$ & $\text{(ATLAS~\cite{Aad:2014wea})}$ \\[0.1cm]
$2\phi+\text{jet}$ & $M\gtrsim 822.8~\text{GeV}$ & $\text{(CMS~\cite{Chatrchyan:2013mys})}$~.\\
\end{tabular}
}
\end{equation}
While these findings are in agreement with the dark matter searches~\cite{cmsjets} recast in~\cite{Brax:2015hma}, we note that the cut scenarios devised in searches for Supersymmetry~\cite{Aad:2014wea,Chatrchyan:2013mys} are slightly better tailored towards dark energy scalar searches. This already sheds some light on the possible discrimination of the nature of a dark energy signature from dark matter signatures. We will discuss this further below.

The limits on ${\cal{L}}_2$ from $t\bar t+p_{T,\text{mis}}$ searches are
\begin{equation}
\label{eq:liml2t}
\parbox{0.37\textwidth}{
\begin{tabular}{c >{\hspace{0.3cm}} l >{\hspace{0.3cm}} r }
${\cal{L}}_2 $ & $M\gtrsim 461.2~\text{GeV}$ & $\text{(ATLAS~\cite{Aad:2014wea})}$ \\[0.1cm]
$2\phi+t\bar t$ & $M\gtrsim 399.8~\text{GeV}$ & $\text{(CMS~\cite{Chatrchyan:2013mys})}$~.\\
\end{tabular}
}
\end{equation}
As expected these limits are not as strong as the ones that are obtained from mono-jet signatures, as large momentum transfer configurations in $t\bar t+p_{T,\text{mis}}$ have a smaller differential cross section, leading to a decreased sensitivity of top pair and missing energy searches compared to mono-jet analyses.

Together, the results of Eqs.~\eqref{eq:liml1j}-\eqref{eq:liml2t} allow us to draw the conclusion that the leading dark energy interactions can be constrained by combining $t\bar t$ and mono-jet searches, with current constraints ranging in the few hundred GeV regime, based on the LHC run I analyses provided in {\sc{CheckMate}}. These constraints can be expected to be pushed during run II (100/fb) with further improvements possible during the LHC high luminosity phase. They provide important  complementary information to other existing searches for dark energy and we encourage the experimental community to perform missing energy searches as outlined above also in the dark energy context.

\subsection{Comparison with LHC dark matter phenomenology}
\label{sec:darkmatter}
A question that becomes important in case of a missing energy-related new physics discovery at the LHC is pinning down, or excluding its relation to dark energy. In case of Supersymmetry, we can expect new exotic states to accompany a missing energy signature in complementary searches, while in dark matter scenarios, similar to dark energy, additional degrees of freedom can lie beyond the kinematic coverage of LHC searches~\cite{Goodman:2010yf,Goodman:2010ku,Fox:2011pm,Buchmueller:2013dya,Abdallah:2014hon,Abdallah:2015ter}. This prompts us to the question: can we tell a difference between the leading dark energy interactions and a similar scalar dark matter scenario? To this end, we show in Fig.~\ref{fig:dmcomp} the normalised expected $p_{T,\text{mis}}$ distributions of the mono-jet and $t\bar t +p_{T,\text{mis}}$ channels for ${\cal{L}}_1$ and ${\cal{L}}_2$ alongside the $p_{T,\text{mis}}$ spectrum of a simplified dark matter model characterised by
\begin{equation}
	{\cal{L}}_{\text{BSM}} \supset {\cal{L}}_{\text{SM}} + {1\over 2} g_\phi\phi^2 Y + {1\over \sqrt{2}} \sum g_i \bar f_i f_i Y,
\end{equation}
where we consider a scalar mediator $Y$ coupling to SM fermions $f_i$.  We set the mediator mass $m_Y =$1 TeV, such that our comparison is not affected by $Y$ going on-shell.

As can be seen from Fig.~\ref{fig:dmcomp}, the energy dependence of a typical dark matter scenario (motivated through a Higgs portal interaction for instance) differs from the dark energy scalar production. While dark energy signatures can be constrained by adapting dark matter searches, their phenomenology is intrinsically different. This provides a new avenue to look for physics beyond the Standard Model through analyses that are specifically tailored to dark energy signals, which will likely result in a better sensitivity than quoted in Eqs.~\eqref{eq:liml1j}-\eqref{eq:liml2t}.

\begin{figure}[!t]
  \centering
  \includegraphics[height=6.0cm]{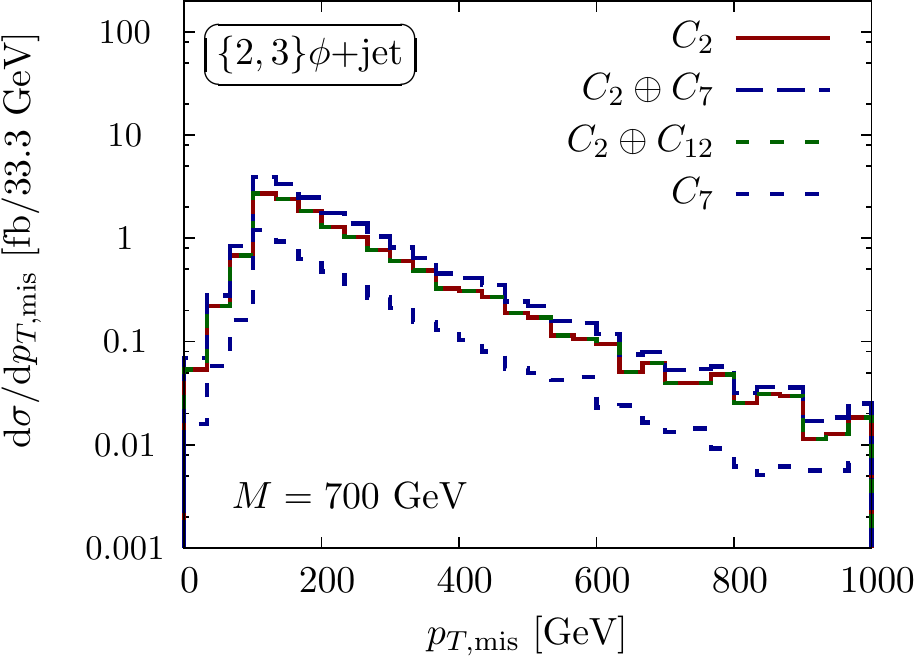}
  \caption{\label{jet:c2nonlin} Missing transverse momentum distribution for the mono-jet channel with $\phi$ multiplicities up to three.}
\end{figure}
\begin{figure}[!t]
  \centering
  \includegraphics[height=6.0cm]{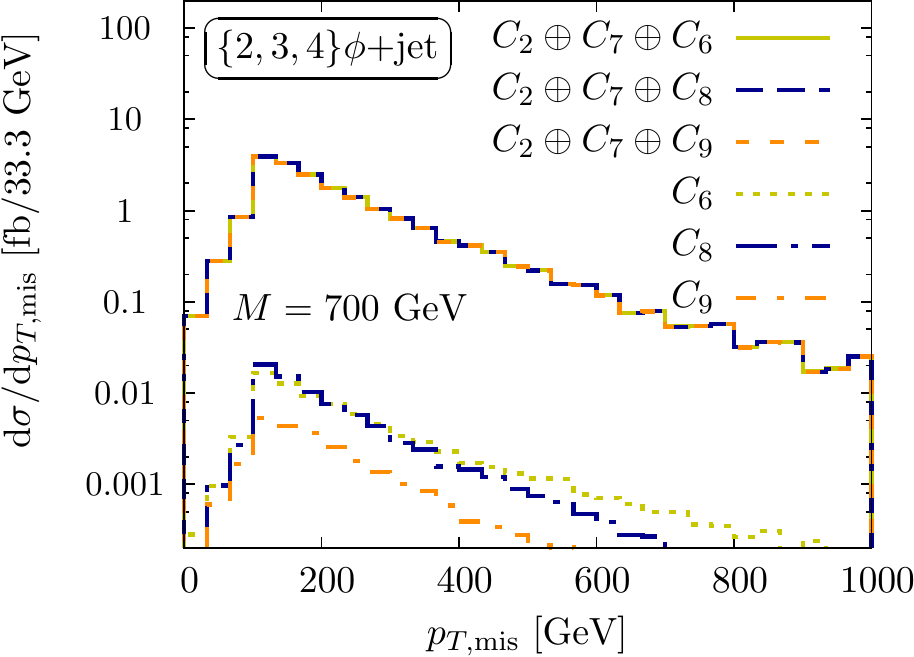}
  \caption{\label{jet:c2nonlinc7} Missing transverse momentum distribution for the mono-jet channel with $\phi$ multiplicities up to four, based on combining $C_2$ with $C_7$. The distributions of $C_{6,8,9}$ are shown separately for comparison.}
\end{figure}
\begin{figure}[!t]
  \centering
  \includegraphics[height=6.0cm]{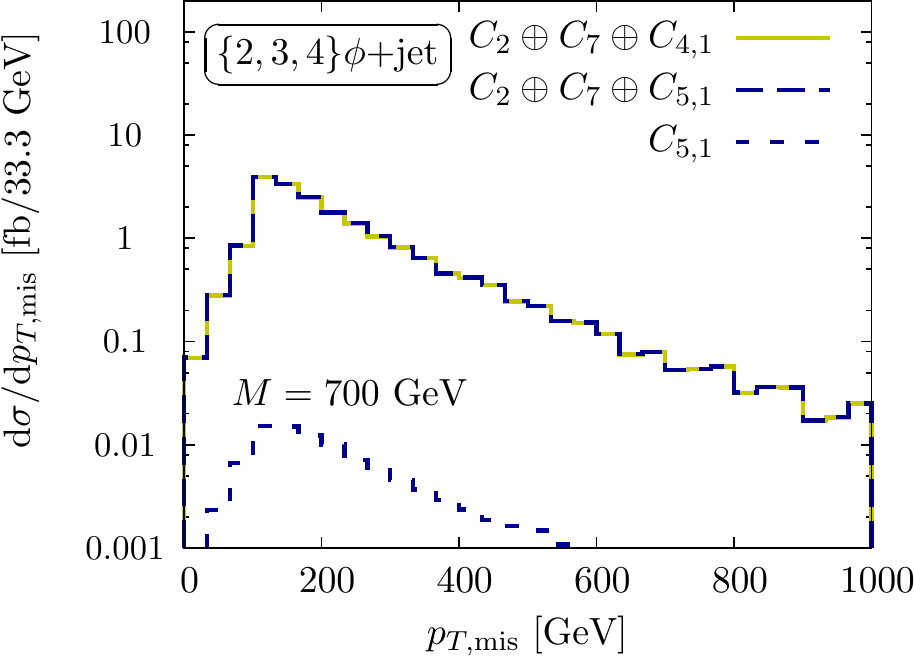}
  \caption{\label{jet:c2nonlinc3a} Same as Fig.~\ref{jet:c2nonlinc7}, however considering the interactions arising from the higher order terms $C_{4,1}$ and $C_{5_1}$. The impact of $C_{3,1}$ is even more suppressed than $C_{4,1}$ and we do not include it to the histogram.}
\end{figure}

\subsection{Phenomenological tests of higher order operators}
\label{sec:collb}
So far we have limited ourselves to the operators ${\cal{L}}_{1,2}$, i.e. the leading interactions of scalar dark energy with the SM sector discussed in Sec.~\ref{sec:eff}, i.e. ${\cal{L}}_{i,1}, 3\leq i\leq 5$ and ${\cal{L}}_{12,4,3}$ (focussing on standard propagators). Given the intrinsic non-linear structure of scalar dark energy, it is worthwhile to address the question of whether these interactions impact the limit setting. Alternatively, if they turn out to have a significant impact (i.e. for a comparably low $M$) we might be able to use collider measurements to formulate a refined picture of the dark energy nature.

The phenomenology of the higher order operators introduced in Sec.~\ref{sec:eff} can be classified according to the dark energy scalar multiplicity in the final states. Since they all lead to the same signature, i.e. they contribute to missing energy, we may add the respective $\phi$ multiplicities incoherently to the full hadronic final state to include the effects of the higher order operators. The number of $\phi$ fields in a particular operator dictates the number of effective operator insertions, which again determines the effective scaling of a cross section with the scale $M$. For instance, ${\cal{L}}_7$ describes a scalar self-interaction and will not contribute to $2\phi$ production with for our case $C_{10}=0$. However it can be combined with ${\cal{L}}_{1,2}$ to obtain a $3\phi$ final state with a scaling $\sim M^{-7}$ at the amplitude level. Note, that this way the interactions ${\cal{L}}_{1,2}$ are probed by one additional off-shell leg and probe the operators ${\cal{L}}_{1,2}$ in a different way. Again we set the Wilson coefficients $C_i=1$ in the following.

\begin{figure*}[!t]
  \centering
   \centering
  \subfigure[\label{top:c1nonlin}]{\includegraphics[height=6.0cm]{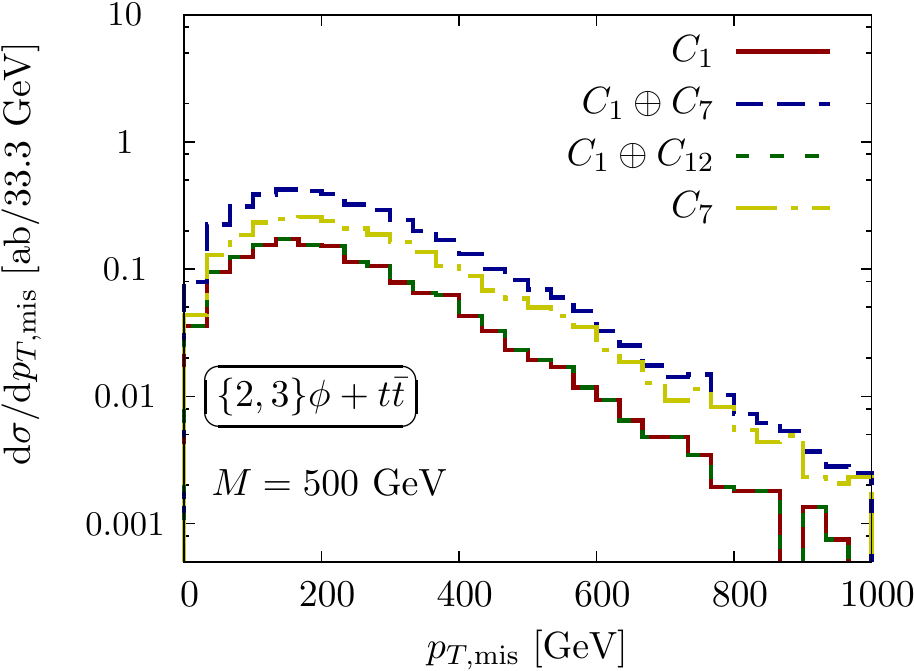}}
  \hfill
  \subfigure[\label{top:c2nonlin}]{\includegraphics[height=6.0cm]{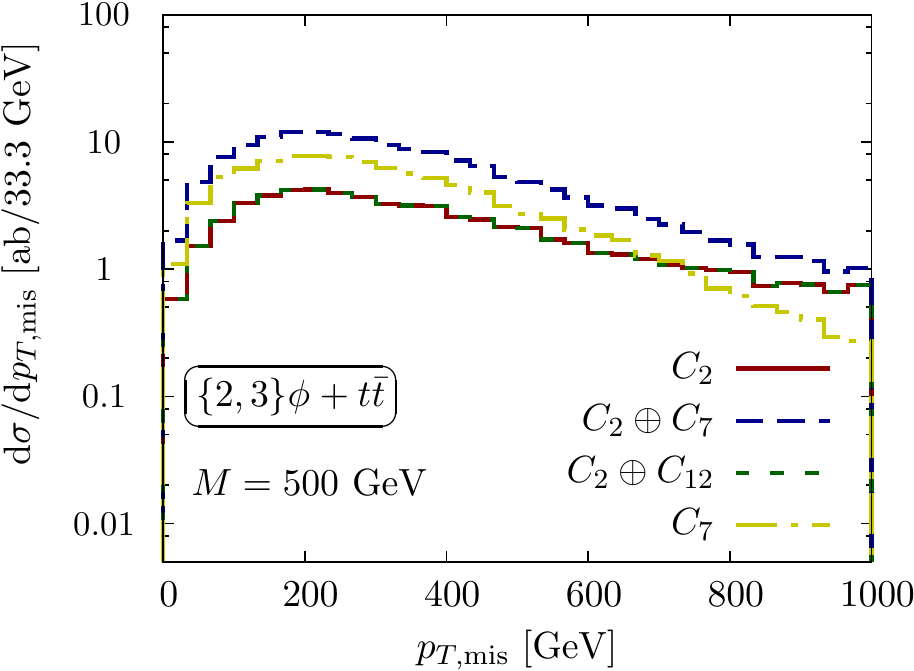}}
  \caption{\label{top:nonlin} Same as Fig.~\ref{jet:c2nonlin} expect that we consider the interactions parameterised by $C_1$ (a) and $C_2$ (b) for the $t\bar t+p_{T,\text{mis}}$ final state.}
\end{figure*}

\begin{figure*}[!t]
  \centering
   \centering
  \subfigure[\label{top:c1nonlin2}]{\includegraphics[height=6.0cm]{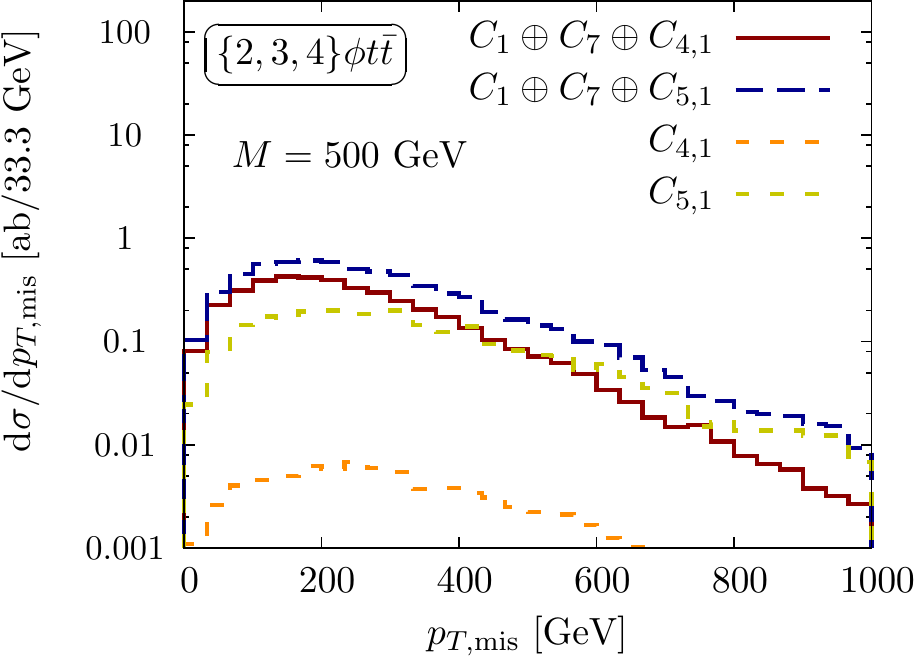}}
  \hfill
  \subfigure[\label{top:c2nonlin2}]{\includegraphics[height=6.0cm]{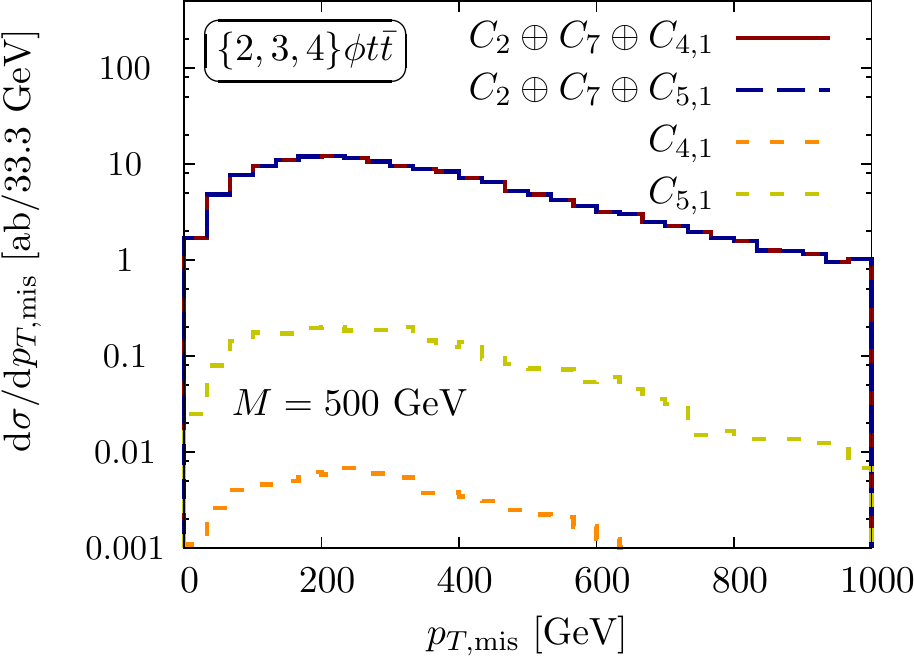}}
  \caption{\label{top:nonlin2} Same as Fig.~\ref{jet:c2nonlinc3a} expect that we consider the interactions parameterised by $C_1$ (a) and $C_2$ (b) for the $t\bar t+p_{T,\text{mis}}$ final state for the operators $C_{4,1}$ and $C_{5,1}$.}
\end{figure*}

Starting from Fig.~\ref{jet:c2nonlin}, we show the effects of combining different operators and $\phi$ multiplicities up to four in Fig.~\ref{jet:c2nonlinc3a} for the operator $C_2$, for which we can formulate constraints in the first place. We choose $M=700~\text{GeV}$ inspired by our results of the previous section. From Fig.~\ref{jet:c2nonlin}, it becomes apparent that the only operator that significantly adds $3\phi$ in comparison to $2\phi$ production is $C_7$, while $C_{12}$ has a negligible effect. In general, for $4\phi$ production, while the energy dependence amongst the different operators $C_{3,1},C_{4,1},C_{5,1},C_{6},C_{8},C_{9}$ is different (see in particular Fig.~\ref{jet:c2nonlinc7}), their overall contribution in light of the constraints obtained in Sec.~\ref{sec:colla} is negligible.

We repeat the same analysis for the $t\bar t+p_{T,\text{mis}}$ channel in Figs.~\ref{top:nonlin} and \ref{top:nonlin2}, with scale choice $M=500~\text{GeV}$ following our discussion in Sec.~\ref{sec:colla}. For comparability, we also choose the same $M$ for the limits from $C_1$, although the current constraints on $M$ are considerably lower.

The qualitative impact of the higher-order interactions is analogous to the mono-jet channel and the comparison of $t\bar t+p_{T,\text{mis}}$ with jet$+p_{T,\text{mis}}$ shows the higher order operator's qualitative behavior as a function of $M$. As we have adopted a lower scale than in the jet+$p_{T,\text{mis}}$ channel we see that operators like ${\cal{L}}_{4,1},{\cal{L}}_{5,1}$ that share similarities with ${\cal{L}}_{2}$ in terms of their structure of $\phi$-derivatives start to compete with the $2\phi$ final state, Fig.~\ref{top:c1nonlin2}. When limits on $M$ are weak, this can mean that the higher order operators can dominate the phenomenology of a particular missing energy search. Such a result needs to interpreted with care as it might correspond to a breakdown of perturbation theory. In the particular case of ${\cal{L}}_1$, however, the tree level effects can be suppressed by requiring a relatively small explicit violation of conformal invariance, while the effects of e.g. ${\cal{L}}_5$ are not restricted by an approximate chiral invariance of ${\cal{L}}_{\text{SM}}$ (this lead to stronger constraints on ${\cal{L}}_2$ in Sec.~\ref{sec:colla}) and mediate prompt $4\phi$ production. If operators fall into the same category, however, such as ${\cal{L}}_{4,n}$ and ${\cal{L}}_{5,n}$, competing multiplicities signal a poor convergence of the effective theory. For example, by choosing different $\phi$ multiplicities to obtain different loop orders contributing to, say, the top 2-point function, we can see that different loop orders start to become equally important, influencing the top lifetime which is related to the imaginary part of the 2-point function.

While the relative size of the operators depends on a particular scalar dark energy scenario,better adapted search strategies as well as increased statistics will push the scale also for these interactions to $\sim 700~\text{GeV}$, which effectively restores a good behavior in the multiplicity scaling pattern that we already observe for ${\cal{L}}_{2}$ in Fig.~\ref{top:c2nonlin2}. In this case ${\cal{L}}_{7}$ is the only interaction that still leaves a sizable impact, and can then be constrained if a new physics discovery exhibits a dark energy character.

\subsection{On the phenomenology of shift-symmetry breaking theories}
\label{sec:shift}
Our analysis so far is valid for coefficient choices $C_{10}$ that leave the scalar stable on collider distances. If there is a significant $C_{10}\neq 0$, the phenomenology dramatically changes as the scalar can be singly-produced and can decay to lighter hadrons, leptons, or photons. For example, the operator ${\cal{L}}_{10}$ introduces interaction vertices with fermions $f_{i}$ (where the index describes the fermion generation) of the form
\begin{equation}
\label{eq:decay}
\parbox{2.2cm}{\includegraphics[scale=0.8]{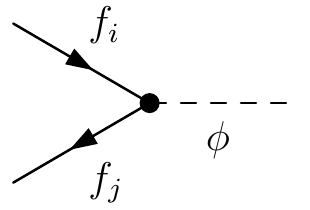}}\quad = {4iC_{10}\over N} m_{f_i} \delta_{ji}\,.
\end{equation}
Depending on the scenario, this can lead to spectacular signatures that range from (highly) displaced vertices similar scenarios of hidden valley or Supersymmetry (see~\cite{Tarem:2005gxa,Ciapetti:2008zz,Nisati:330302,Ambrosanio:468069,Ellis:1006573}) to emerging signatures in the different layers of the detector~\cite{Englert:2011us,Schwaller:2015gea}. These signatures are fundamentally different from the ones that we have discussed so far and a comprehensive investigation is beyond the scope of this work.\footnote{It is however worthwhile to remark that since ${\cal{L}}_{10,1}$ effectively describes an interaction of a Higgs boson (i.e. it couples like a pseudo-dilaton), the dark energy phenomenology shares the signatures of light Higgs portal scalars discussed in~\cite{Englert:2011us,Curtin:2013fra} as well as on-going efforts within the Higgs Cross Section Working Group~\cite{YR4}. This includes in particular the phenomenology of heavy $\phi$ bosons in the TeV range.}. Eq.~\eqref{eq:decay} leads to a partial $\phi$ decay width into fermions
\begin{equation}
\label{eq:decwidth}
	\Gamma(\phi\to f\bar f) = {2\over \pi} {C_{10}^2} {m_f^2\over N^2} {(m_\phi^2-4m_f^2)^{3/2}\over m_\phi^2}\, ,
\end{equation}
which leads to a traveled distance through the detector
\begin{equation}
D={\beta\gamma	\over \Gamma_\phi}
\end{equation}
where $\Gamma_\phi$ is the total decay width, which is dominated by the size of the effective Yukawa interaction $\sim C_{10} m_f/N$ if sufficient phase space is available, i.e. for $\phi$ masses not too close to the respective decay threshold. The width is typically very small, and for a mass of $0.1~\text{GeV}$ we obtain a total decay width of $\sim 2\times 10^{-10}$~GeV.

The probability of decaying between distances $L_1 < L_2$ is then given by
\begin{equation}
P(L_1\leq L \leq L_2 ) =\int_{L_1}^{L_2} \hbox{d}L'\, {1\over D} \exp\left( - {{L' \over D}}\right)\,.
\end{equation}

To get an idea of the resulting phenomenology, we consider a dark energy scalar with mass $m_\phi=20~\text{GeV}$ and its decay $\phi \to b\bar b$ produced at $p_T\simeq 100~\text{GeV}$ (the typical scale of a mono-jet configuration). For this mass choice the decay is open and enhanced over the other channels. For a choice $C_{10}m_b/N \simeq 10^{-8}$ we can expect that around 99\% of the produced $\phi$ bosons will decay inside the detector $\lesssim 7~\text{m}$ (using the transverse CMS dimensions in this particular case): 54\% of decays in the tracker ($L\lesssim 1~\text{m}$), 41\% inside the electromagnetic and hadronic calorimeters ($1~\text{m}\lesssim L\lesssim 4~\text{m}$) and 4\% inside the muon detectors ($4~\text{m}\lesssim L\lesssim 7~\text{m}$).\footnote{Due to a different geometry, we can expect a slightly better coverage by the ATLAS experiment.} The search strategies in each part of the detector depends on trigger and selection criteria as well as on the calibrated performance of each part of the detector. For instance, fermions are typically stripped off in the first layers of the muon system, hence a decay $\phi\to b\bar b$ in that region of the detector would be considered as noise. On the other hand, decays inside the tracker whose high resolution enables the search for displaced vertices makes this part of the parameter space accessible.

The phenomenology strongly depends on the effective and dominant Yukawa interaction $C_{10} m_b/N$. Increasing $C_{10} m_b/N\simeq 5\times 10^{-8}$ all particles decay inside the tracker with $99\%$ of $p_T\simeq 100~\text{GeV}$ events decay with displaced vertices, whereas for $C_{10} m_b/N\simeq 10^{-6}$ the $\phi$ bosons will decay before leaving a displaced vertex signal. In such a case, additional reconstruction techniques are available but subject to detector systematics as well as large QCD backgrounds.

The considerably larger scales that can be probed  with displaced vertex searches (note that this also applies to different quark flavors and leptons other than the bottom considered in this example) should allow to probe scales in the region $N\sim 10^8~\text{GeV}$, which will provide comparably stronger constraints on $N$ than on $M$. For the latter we lose sensitivity as soon as the scalar is allowed to decay inside the detector\footnote{Note that even for $\phi$ masses below the lightest fermion thresholds, the loop-induced decay to photons and gluons can still dominate.}. This provides an interesting and complementary avenue to look for dark energy scalars on the basis of existing searches. We leave a more detailed investigation to future work~\cite{future}.

\section{Summary And Conclusions}
\label{sec:conclusions}
The mystery of dark energy is motivation to consider new physics that is relevant on cosmological scales. In particular the possibility that light dark energy scalar fields might exist and interact with the Standard Model. In this paper we have surveyed a large class of effective dark energy interactions and motivated the combination of mono-jet and $t\bar t+p_{T,\text{mis}}$ analyses to constrain the leading aspects of dark energy interactions with the SM sector at the LHC. In passing we have used the phenomenological signatures in these channels to obtain the latest LHC constraints on the dominant dark energy signatures by recasting existing 8 TeV Supersymmetry and dark matter analyses. While dark energy signatures share some aspects of dark matter phenomenology, the dark energy signatures are in general different, and provide a new phenomenological avenue to look for well-motivated signs of physics beyond the SM. In case a new physics discovery is made that falls in to the category of a scalar dark energy signal, some aspects of the dark energy scalar's self-interactions can be probed by investigating the missing energy-dependence of the new physics signal, depending on the particular dark energy model. In particular, the discrimination from the competing scalar dark matter interpretation will become possible. Allowing the presence of shift symmetry-breaking operators, the sensitivity to shift symmetry-conserving operators is decreased when scalar decays on collider scales becomes possible. In such a case searches for displaced vertices provide an avenue to constrain the presence of such scalars for relatively large scales.

\medskip

\noindent{\it{Acknowledgements}} --- PB acknowledges partial support from the European Union FP7 ITN INVISIBLES (Marie Curie Actions, PITN- GA-2011- 289442). MS is supported in part by the European Commission through the ``HiggsTools'' Inital Training Network PITN-GA-2012-316704. CB is supported by a Royal Society University Research Fellowship and in part by the IPPP Associateship programme.  We would like to thank Eugene Lim, Andrew Pilkington and David Seery for very helpful discussions in the preparation of this work.


\bibliography{paperfinal}

\end{document}